\documentclass[prd,fleqn,twocolumn]{revtex4-2}

\usepackage[english]{babel}
\usepackage{graphicx}

\usepackage{mathtools}
\usepackage{amsfonts}
\usepackage{amsmath}
\usepackage[utf8]{inputenc}
\usepackage{xcolor}
\usepackage{physics}
\usepackage{overpic,color}
\usepackage{hyperref}
\usepackage{dsfont}
\usepackage[charter,expert]{mathdesign}
\usepackage[normalem]{ulem}
\usepackage{cancel}
\usepackage[mathscr]{euscript}

\newcommand{\hamiltonianglobal}{\hat{H}_\mathrm{tot}}
\newcommand{\hamiltoniansystem}{\hat{H}}
\newcommand{\hamiltonianclock}{\hat{H}_{\mathscr{E}}}
\newcommand{\hamiltonianenv}{\hat{h}_{\mathscr{E}}}
\newcommand{\energyglobal}{E_\mathrm{tot}}

\newcommand{\interactionfunction}{V}
\newcommand{\interaction}{\hat{\interactionfunction}}

\newcommand{\stateglobal}{\Psi}

\newcommand{\prm}{\lambda}
\newcommand{\prs}{\gamma}
\newcommand{\prg}{\Lambda}

\newcommand{\ketsystem}[1]{\ket{#1}}
\newcommand{\ketclock}[1]{\ket*{#1}_{\mathscr{E}}}
\newcommand{\ketglobal}[1]{\ket*{#1\rangle\!}}

\newcommand{\braketclockglobal}[2]{\braket{#1}{#2\rangle\!}}

\newcommand{\melclock}[3]{\mel{#1}{#2}{#3}_{\mathscr{E}}}

\newcommand{\melglobal}[3]{\langle\!\langle #1 | #2 | #3 \rangle \!\rangle}

\newcommand{\ident}{\hat{\mathds{1}}}
\newcommand{\identsystem}{\ident}
\newcommand{\identclock}{\ident_{\mathscr{E}}}
\newcommand{\dyadclock}[1]{\dyad{#1}_{\mathscr{E}}}
\newcommand{\dyadglobal}[2]{\dyad*{#1\rangle\!}{\!\langle#2}}

\newcommand{\projector}{\hat{P}}
\newcommand{\projectorglobal}{\hat{P}_{\stateglobal}}

\newcommand{\expvalsystem}[1]{\langle #1 \rangle}
\newcommand{\expvalclock}[1]{\langle #1 \rangle_{\mathscr{E}}}

\newcommand{\trsystem}{\tr}
\newcommand{\trclock}{\tr_{\mathscr{E}}}

\newcommand{\densitysystem}{\hat{\rho}}

\newcommand{\densityclock}{\hat{\rho}_{\mathscr{E}}}

\newcommand{\partitionfunc}{Z}

\newcommand{\partitionfuncsystem}{{\partitionfunc}}

\newcommand{\jmx}[1]{\textcolor{black}{#1}}
\newcommand{\jmr}[1]{\textcolor{black}{#1}}
\newcommand{\seb}[1]{\textcolor{black}{#1}}

\begin{document}

\title{Dynamics of temperature from a relational approach}
\title{Statistical mechanics from relational complex time with a pure state}
\author{Sebastian Gemsheim}
\email{sebgem@protonmail.com}
\author{Jan M.~Rost}
\affiliation{Max Planck Institute for the Physics of Complex Systems, N\"othnitzer Stra\ss e 38, D-01187 Dresden, Germany}

\begin{abstract}
Thermodynamics and its quantum counterpart are traditionally described with statistical ensembles. Canonical typicality
has related statistical  mechanics for a system to ensembles of global energy eigenstates of system and its environment analyzing their cardinality.  We show that the canonical density for a system emerges from a maximally entangled global state of system and environment through relational complex time evolution between system and environment without the need to maximize the entropy or to count states.
\end{abstract}

\maketitle

%%%%%%%%%%%%%%%%%%%%%%%%%%%%%%%%%%%%%%%%%%%%%%%%%%%%%%% BODY
Canonical typicality has proven to be a powerful concept to understand how statistical mechanics emerges for a (small) system $\mathscr S$ singled out from its environment $\mathscr E$ where environment and system together are in a  pure energy eigenstate of the global Hilbert space \cite{gole06, Popescu2006}. This is achieved by analyzing
what is "typical" about the ensemble of  excessively many possibilities to realize such an eigenstate in a small energy interval $[E,E+\delta E]$ and assessing the exponentially growing number of typical realizations in ensembles of  increasing energy $E$. The approach  carries the spirit of statistical mechanics which deals with many degrees of freedom and therefore naturally with ensembles. Yet,  quantum mechanics  introduces already for a single degree of freedom a probabilistic perspective with the wavefunction associated to its dynamics. Hence, one could hope to capitalize on this fundamental quantum property to derive statistical mechanics from a single pure state. The fact that, compared to classical mechanics, quantum mechanics offers with entanglement another resource for variety already encoded in a single state renders the existence of such a state as the source of statistical mechanics more likely. 
 We will show that statistical mechanics is contained in a single, {\it maximally entangled} global eigenstate (MES)  of system and environment without the need to consider ensembles of global states at different energies \cite{gole06,Popescu2006} or to count degenerate states of the system \cite{dezu16}.  Instead, we extend the relational time approach originally conceived by Page \& Wootters \cite{Page1983} to propagate density matrices in imaginary relational time. This will allow us to conceive the fundamental thermodynamic relation from the density matrix of  a maximally entangled global state for any desired imaginary time $\gamma$, which will turn out to represent temperature in the well known way, $(k_BT)^{-1} = \gamma$,
 if infinite temperature $\gamma = 0$ is linked to the MES. Hence, relational complex time evolution turns out to be the element which completes the quest for deriving statistical mechanics from a pure quantum state without the need to resort to additional statistical considerations.
 The result confirms the conjecture that temperature, similarly as time, has a relational root \cite{Vedral2017} and provides a physical justification for the striking connection between dynamics and thermal physics~\cite{Feynman2010}.

Relational time emerges from a splitting of the global Hamiltonian $ \hamiltonianglobal 
    {=} \hamiltoniansystem \otimes \identclock + \identsystem \otimes \hamiltonianclock + \interaction$ into system $\hamiltoniansystem$ and environment $\hamiltonianclock$ 
 interacting with each other through $\interaction$.   
Since the global state $\stateglobal$ fullfills $(\hamiltonianglobal - \energyglobal) \ketglobal{\stateglobal} = 0$   one can formulate the  \emph{invariance principle} \cite{Gemsheim2023a}
\begin{equation}
    e^{\prg(\hamiltonianglobal - \energyglobal)} \ketglobal{\stateglobal} 
    = \ketglobal{\stateglobal}
    \label{eq:invariance_complex_time}
\end{equation}
which does not only hold for all imaginary-valued symmetry parameters $\prg = i\prm$,  but also for real ones $\prg = \prs$, or in general for complex $\prg$. 

With \eqref{eq:invariance_complex_time} we can write for the density matrix   $P_\stateglobal=\dyadglobal{\stateglobal}{\stateglobal}$ of the global state
\begin{equation}
    \projectorglobal=  e^{\prg(\hamiltonianglobal - \energyglobal)} \projectorglobal e^{\prg^*(\hamiltonianglobal - \energyglobal)}\,.
    \label{eq:invariance_dm}
\end{equation}
From now on we will assume in accordance with the original proposal for relational time~\cite{Page1983} that the interaction $V$ is small enough to be neglected. In fact, this condition can be softened without loosing the thrust of the relational time concept by only demanding that $[\hamiltonianclock,\interaction]=0$, 
implying that the environment's eigenstates are not changed by the interaction, as we will see later.

For now with $\interaction {=} 0$, the evolution according to the Hamiltonians for the
environment  $\hamiltonianenv \equiv \hamiltonianclock-\energyglobal$ and system $\hamiltoniansystem$ in \eqref{eq:invariance_dm} factorizes, 
\jmr{
\begin{equation}
\label{eq:factorization}
    e^{\prg(\hamiltonianglobal-\energyglobal)}=e^{\prg\hamiltonianenv}\jmr\otimes e^{\prg\hamiltoniansystem}\,,
\end{equation}
}
and we can obtain from \eqref{eq:invariance_dm} the density matrix of the system evolved in complex time $\prg$,
\begin{subequations}
\label{eq:relational}
\begin{equation}
    \densitysystem(\prg) =  \frac{1}{\partitionfuncsystem(\prg)}e^{-\prg\hamiltoniansystem} \densitysystem_0 e^{-\prg^*\hamiltoniansystem}\,,
     \label{eq:dm_system}
\end{equation}
where
\begin{equation}
e^{-\prg\hamiltoniansystem}\densitysystem_0 e^{-\prg^*\hamiltoniansystem} =  \trclock[\densityclock(-\prg^*) \projectorglobal]
  \label{eq:dm_env}
\end{equation}
with
\begin{equation}
    \densityclock(\prg) = \, e^{-\prg\hamiltonianenv} \densityclock(0) e^{-\prg^*\hamiltonianenv}\,,
    \label{eq:invariance_d_system}
\end{equation}
\end{subequations}
which follows from \eqref{eq:invariance_dm}
after tracing  over the environment. We have left room for a conditioning density  $\densityclock(0)$ of the environment, such that $\densitysystem(0)=\partitionfuncsystem(0)^{-1}\trclock [\densityclock(0) P_\stateglobal]$.
 The normalization $\partitionfuncsystem(\prg)=\melglobal{\stateglobal}{\densityclock(\jmr{-\prg^*})}{\stateglobal}$ ensures
that $\trsystem\densitysystem(\prg)=1$.

Note that according to \eqref{eq:dm_env} the evolution of the system density $\densitysystem$ with $\prg$ means evolution of the environment density with $-\prg^*$ illustrating the relational character of the evolution in system and environment, respectively.
For real and continuous \jmr{$\prg = \gamma$}, differentiation of \eqref{eq:dm_system}  leads to
\begin{equation}
    d \densitysystem(\prs) /d\prs
    = - 
    \Bigl\{ \hamiltoniansystem - \expvalsystem{\hamiltoniansystem}(\prs)  , \densitysystem(\prs) \Bigl\} \,,
    \label{eq:imaginary_time_von_Neumann_equation}
\end{equation}
the von Neumann equation for imaginary time $\prs$ emerging from complex relational time.  The  mean energy $\expvalsystem{\hamiltoniansystem} = \trsystem(\hamiltoniansystem \densitysystem)$ of the system  in the anti-commutator $\{ \cdot, \cdot \} $
 ensures the normalization. 
For real time $\prg=i\prm$, $\prm\in \mathds{R}$, the standard von Neumann equation
\begin{equation}
d\densitysystem(\seb{i\lambda})/d\lambda = -i[\hamiltoniansystem,\densitysystem(\seb{i\lambda})]\,,
    \label{eq:dm_system_lambda}
\end{equation}
 for the time-dependence of the density matrix follows from
 \eqref{eq:dm_system}  by differentiation. The conditioning environmental density $\densityclock$
in $\densitysystem_0=\trclock [\densityclock(0) P_\stateglobal]$ allows one to specify different initial conditions with the
global state $\ketglobal{\stateglobal}$ as discussed for wavefunctions in \cite{Gemsheim2023a}. Both differential equations hold of course for all global states satisfying \eqref{eq:invariance_complex_time}. 

Statistical mechanics is encoded in \eqref{eq:relational}. Standard tracing over the environment  (\jmr{implying} $\densityclock(0) \propto \identclock$) with eigenstates $\ketclock{J}$ \jmr{and eigenenergies $E_J$}
of $\hamiltonianclock$ \jmr{in \eqref{eq:dm_env}}
renders the system density diagonal. \jmr{Denoting $\ketsystem{\phi_J}=\braketclockglobal{J}{\stateglobal}$ and}  $\prg = \prs/2 + i\prm$ we get from \jmr{\eqref{eq:dm_env}
using  \eqref{eq:invariance_d_system}}
\begin{align}
\label{eq:density_canonical}
   \densitysystem(\prg)
   &=\frac{1}{\partitionfuncsystem(\prg)}\sum_J \melclock{J}{e^{\prg\hamiltonianenv} \projectorglobal e^{\prg^*\hamiltonianenv} }{J}  \nonumber \\ 
   & =\frac{1}{\partitionfuncsystem(\prg)}\sum_{\jmr{J}} e^{-\prs\epsilon_J}\dyad{\phi_J}\\
   &= \frac{1}{\partitionfuncsystem(\prg)}e^{-\prs\hamiltoniansystem}\densitysystem_0 \nonumber\,,
\end{align}
\jmr{where $\epsilon_J = E-E_J$ and the normalization $\partitionfuncsystem(\prg)$ absorbs
the constant of proportionality in $\densityclock(0) \propto \identclock$. The last identity follows from the invariance principle and \eqref{eq:factorization}.}

\jmr{Obviously, \eqref{eq:density_canonical}} represents the canonical density for all temperatures if one identifies $\gamma = 1/(kT)$  and if $\braket{\phi_J}{\phi_{J'}} = c\delta_{JJ'}$,
where the constant $c$ once again can be absorbed in $Z(\prg)$ and therefore may be set to unity, $c=1$. Then, evaluating the
 entropy 
 \seb{$S = - \trsystem (\densitysystem \ln \densitysystem)$}
 with \eqref{eq:density_canonical} gives $S(\jmr{\prs/2}) =  \prs\expvalsystem{\hamiltoniansystem}(\jmr{\prs/2}) + \ln\partitionfuncsystem(\jmr{\prs/2})$  implying the thermodynamic relation $dS = \prs d \expvalsystem{\hamiltoniansystem}$. 

\jmr{Note that we have derived the canonical density for the system  solely by the relational connection to the global state apart from requiring the $\ket{\phi_J}$ to be orthonormal. This requirement has a physical reason:}
With a system at infinite temperature $\gamma {=} 0$ we associate  minimal structure implying in the relational context that  the system states $\ketsystem{\phi_J}$ with their different energies $\epsilon_J$ should be contained in the global state with equal probability $\braket{\phi_J}{\phi_J}=
 \melglobal{\Psi}{J\rangle\langle J}{\Psi} = 1$ for all $J$. The global state with this property is maximally entangled,
\begin{equation}
    \label{eq:max_globalstate}
    \ketglobal{\stateglobal_M}= \sum_J \jmr{ \ketclock{J} \otimes \braketclockglobal{J}{\stateglobal_M} } = \sum_J \ketclock{J}\otimes\ketsystem{\phi_J}\,,
\end{equation}
where  the $\{\ket{\phi_J}\}$ form an \jmr{orthonormal basis} for the system Hamiltonian with eigenenergies
$\epsilon_J$. 
Note, that thanks to the MES and with
 $\expvalclock{\hamiltonianclock} = \trclock(\hamiltonianclock\densityclock)$, 
 \begin{equation}
     E = \expvalsystem{\hamiltoniansystem}(\prg)+\expvalclock{\hamiltonianclock}(\prg)
 \end{equation}
holds for all complex times $\prg$,
in particular for all temperatures. The link of the MES to infinite temperature through complex relational time  motivates its role for statistical mechanics which is given in \cite{dezu16} on very general grounds of symmetry considerations for a global pure state describing system and environment.

 This remarkably direct quantum route to the fundamental relation of thermodynamics through relational imaginary time propagation of a maximally entangled global state bears some interesting properties  of  relational complex time,
 we would like to point out in the following. 
\begin{itemize}
\item[(A)] The energy constraint \eqref{eq:invariance_complex_time}
has the consequence, that 
\begin{equation}
     [\densityclock,\hamiltonianenv]=0 \Leftrightarrow [\densitysystem,\hamiltoniansystem]=0\,.
\end{equation}
In other words, if the conditioning density $\densityclock$ is (block) diagonal in the energy eigenstates of the environment,  the system's density matrix $\densitysystem$  commutes with the system Hamiltonian, rendering $\densitysystem$ stationary in real time, see \eqref{eq:dm_system_lambda}. As the imaginary time evolution is ruled by the anti-commutator, the system's density changes in imaginary time, but only through the exponential energy factors, the coefficients in the density remain constant. Since simply tracing over the environment implies  $\densityclock(0) \propto \identclock$ diagonal, this is also true for the imaginary time evolution giving rise to statistical mechanics as derived: The evolution relates via the von Neumann equation in imaginary time the system at different temperatures, maintaining equilibrium (that is stationarity in real time), thereby explicating
the "peaceful coexistence of thermal equilibrium and emergence of time" as put forward in \cite{Favalli2022}.
\item[(B)] 
As a pure state the MES carries vanishing entropy $S_\mathrm{tot}{=}0$. 
  \jmr{A density matrix of the system}  singled out from \jmr{the MES} by tracing over the environment,
 $\densitysystem = \jmr{D^{-1}}\trclock [\jmr{\projectorglobal}] =\jmr{D^{-1}}\sum_{J=1}^D\dyad{\jmr{\phi_J}}$,  is maximally mixed with (maximal) entropy, $S = \ln{D}$, 
\jmr{where $D$ is the dimension of the system's Hilbert space.} This property characterizes the maximal deficit of knowledge or degree of in-determinism, which is typically associated with a system's state at infinite temperature. 
 
 The mathematical connection of a maximally determined global MES with a maximally mixed (in-determined) state of the system 
originates from the way, how information about the system is encoded in the MES, namely by exclusively and evenly linking system wave functions to those of the environment. Hence, giving up the link through tracing over the environment leaves minimal information about the system. On the other hand, a global state which is not an MES is in general a Schmidt state $\ketglobal\stateglobal=\sum_J a_J\ketclock{J}\otimes\ketsystem{\phi_J}$, $a_J\in \mathds{R}$.  Links within (sub-)systems, represented by coefficients   $a_J\ne 1$ in  the global state   are preserved upon cutting the links to the environment by tracing over it, and the resulting system state is not maximally mixed. 
\end{itemize}

In an experimental situation the global system designed  may not be in a maximally entangled or close by state, but only in a Schmidt state. In this case it is still possible to mimic  thermodynamics for the system via a suitable conditioning density $\densityclock$ of the environment. 
With $\densityclock(0) = \sum_J \jmx{p_J} \dyadclock{J}$
 one gets explicitly
\begin{align}
    \densitysystem(\prg) = & \frac{1}{\partitionfuncsystem(\jmr{\prg})}e^{-\prs\hamiltoniansystem}\densitysystem(0)\nonumber\\
    = & \frac{1}{\partitionfuncsystem(\jmr{\prg})}\sum_J e^{-\prs \epsilon_J} \jmx{p_J} a_J^2\dyad{\phi_J}\,,
     \label{eq:dm_system_diagonal}
\end{align}
where $\prg = \prs/2 + i\prm$.  A conditioning density $\densityclock$ with $\jmx{p_J} = a_J^{-2}$ gives rise to a canonical density for the system and therefore statistical mechanics. Such a setup is conceivable using cavities and quantum control.
It may be even more interesting to study with this kind of setup deviations from statistical mechanics with densities of the form
of  \eqref{eq:dm_system_diagonal}. Another step of complication would be to include real time dynamics if $\densitysystem$ does not commute with $\hamiltoniansystem$, directly relevant for eigenstate thermalization \cite{DAlessio2016, Deutsch2018}. Of course this complicates, the expression for entropy which is also true for explicit interaction $V\ne0$ in $\hamiltonianglobal$. 

Partially, the effect of interaction is included through the entanglement in the global state: Without interaction the global state could be a simple product state. Preserving the general outcome, an interaction $\interaction$ in \eqref{eq:invariance_complex_time} can be included which commutes with the environmental Hamiltonian, $[\hamiltonianclock,\interaction]=0$, or due to the Jacobi identity \footnote{$[\hamiltonianclock,[\interaction,\projector_J]]+[\interaction,[\projector_J,\hamiltonianclock]]+[\projector_J,[\hamiltonianclock,\interaction]]=0$, where $\projector_J=\dyadclock{J}$.} equivalently
 $[V,\jmr{\dyadclock{J}}]=0$. 
This means that the environment is big enough such that its eigenfunctions are not modified by $\interaction$. The result \eqref{eq:density_canonical} remains the same, only that now  the eigenfunctions $\ket{\phi_J}$ and eigenenergies $\epsilon_J$ fulfill  the Schr\"odinger equation  $(\hamiltoniansystem +\jmr{\melclock{J}{\interaction}{J}} - \epsilon_J)\ketsystem{\phi_J}=0$.

In summary, we have shown from a relational dynamics point of view  how statistical physics of a system is encoded in  maximally entangled states of the system and its environment.  The observation underscores their pivotal role for fundamental quantum principles, as these states are also key in the derivation of Born's rule with envariance \cite{Zurek2003}. Maximally entangled states have the least structure and are evenly distributed over  system and environment, corroborating the situation at infinite temperature.  Without the need of ensemble considerations such as canonical typicality, but by capitalizing on the probabilistic nature of a  quantum state itself, statistical physics for the system at finite temperature  emerges from  propagation of the maximally entangled state in relational complex time.

We thank A. Eisfeld and F. Fritzsch for helpful discussions.


%\bibliography{bib_time_temperature_connection}

\end{document}